\documentclass[12pt,psf,epsf]{article}
\usepackage[centertags]{amsmath}
\usepackage{amssymb}
\usepackage{graphicx}
\usepackage{epsfig}
\usepackage{ulem}
\usepackage[english]{babel}
\usepackage{array}
\usepackage{amsthm}
\usepackage{latexsym}
\usepackage[mathcal]{euscript}
\pdfoutput=1
\usepackage{epsfig}

\usepackage{jheppub}
\usepackage{hyperref}

\usepackage{subfigure}
\usepackage{psfrag}

\usepackage{epsfig}
\usepackage[latin1]{inputenc}
\usepackage{float}
\usepackage{graphicx}
\usepackage{cancel}
\usepackage{mathrsfs}
\usepackage{amssymb}
\usepackage{amsfonts}
\usepackage{amsmath}
\usepackage{slashed}
\usepackage{graphicx}
\usepackage{bm}
\usepackage{color}
\newcommand{\be}{\begin{equation}}
\newcommand{\ee}{\end{equation}}
\newcommand{\bea}{\begin{eqnarray}}
\newcommand{\eea}{\end{eqnarray}}
\newcommand{\bwt}{\begin{widetext}}
\newcommand{\ewt}{\end{widetext}}

\newcommand{\bi}{\begin{itemize}}
\newcommand{\ei}{\end{itemize}}

\usepackage{setspace}

\definecolor{dgreen}{rgb}{0.,0.6,0.}

\usepackage{ulem}

\begin{document}

\title {Exploring Uberholography}

\author{Dmitry S. Ageev}
\affiliation{Department of Mathematical Methods for Quantum Technologies, Steklov Mathematical Institute of Russian Academy of Sciences, Gubkin str. 8, 119991 Moscow, Russia}
\emailAdd{ageev@mi-ras.ru}

\abstract{In this paper, we study the holographic quantum error correcting code properties in  different  boundary fractal-like structures. We construct and explore different examples of the uberholographic bulk reconstruction corresponding to these structures in higher dimensions for Cantor-like sets, thermal states and $T\overline{T}$-deformed conformal field theories. We show how the growth of the system dimension emphasizes the role of the Cantor set, due to the special bound naturally arising in this context.}

\maketitle

\newpage
\section{Introduction}

The study of the entanglement entropy in the context of holographic duality has led to fruitful investigations of the way how different aspects of quantum information including the quantum error correction (QEC) theory fit in the gravitational picture of quantum information \cite{Chen:2021lnq,Rangamani:2016dms,Faulkner:2022mlp}. Extraordinary connection between such dissimilar on the first sight theories like gravity and QEC has attracted significant interest recently. This connection is essentially based on the notion of the entanglement wedge and hypotheses related to it \cite{Hubeny:2007xt,Czech:2012bh,Jafferis:2014lza}. The QEC theory was applied first to resolve the apparent inconsistency in the relation between subregion-subregion duality and some properties of operator algebra in quantum field theory, paving the way to different extensions of QEC way of understanding of holographic duality \cite{Pastawski:2015qua,Almheiri:2014lwa,Dong:2016eik,Harlow:2016vwg,Pastawski:2016qrs}. Assuming the presence of system correctability in AdS/CFT correspondence, different problems arising from a particular logical subalgebras can be formulated as a problem about AdS bulk geometry. In \cite{Pastawski:2016qrs}, the so-called uberholographic proposal has been made. The (sub)linear scaling of the price and the distance of a logical subalgebra corresponding to the specific holographic code leads to the possibility that the logical subalgebra can be supported on the fractal set with the dimension ``lower'' than the boundary dimension. This fractal dimension is expected to be a code universal feature and shows the way how the bulk emerges from the system with lower dimension than it typically occurs in holography (see also recent proposal about the so-called ``wedge holography'' when the system with $d-2$ dimensions gives rise to the $d$-dimensional bulk).

In \cite{Pastawski:2016qrs}, the uberholographic property and the way it works has been demonstrated on the Cantor set defined on a constant time slice of Poincare AdS and the explicit estimate on the code distance has been obtained. This construction has been extended on the case of $AdS_4/CFT_3$ duality and reconstruction of the Sierpinski triangle defined on the boundary \cite{Bao:2022vxc}. In \cite{Chen:2019uhq}, uberholography has been discussed in the context of black hole evaporation. Paper \cite{Gesteau:2020hoz} has studied uberholography for special Hamiltonians motivated by the HAPPY code. In this paper, we study a bunch of explicit examples where different variations on the construction based on the Cantor-like sets can be implemented. One of the main results of \cite{Pastawski:2016qrs} is the estimate of the distance $d\left(\mathcal{A}_{X}\right)$ of logical subalgebra associated with the fractal region $X$ (which is chosen to be a Cantor set) in the form
\be 
d\left(\mathcal{A}_{X}\right) \leq \left(\frac{|R|}{a}\right)^{\alpha}, \,\,\,\,\alpha=\frac{\log 2}{\log (2 / r)},\,\,\,\,r=2\left( \sqrt{2} - 1 \right),
\ee
where $R$ is the size of the set under consideration\footnote{We mean the set from which we constructed the fractal using some iterative procedure.} and $a$ is the cut-off (elementary lattice spacing). 
We generalize this bound to different setups of excited states and different dimensionalities, noticing some universal features corresponding to each case. 
Let us briefly list our results.
\begin{itemize}
    \item We slice the $(d-1)$-dimensional spatial boundary of the Poincare patch $AdS_{d+1}$ by infinite hyperplanes and obtain that $r=3-\sqrt{5}$ for $d=3$ and $r=\sqrt{3}-1$ for $d=4$. We reconfirm that this estimate is universal for $d=3$ taking other slicings into finite size regions --- thin concentric annuli. For larger $d$, we find that $r=2/3$ serves as the bound for $r$, i.e. exactly the value of $r$  corresponding to the Cantor set.
    \item We consider the holographic dual of finite temperature 2d CFT and the thermal fractal. As a result, we find the explicit dependence of $r$ on the combination $\gamma=T\cdot R$ (where $R$ is assumed to be relatively small and temperature $T$ to be large, such that $\gamma$ is finite). The generalization to rotating BTZ black hole is also presented.
    \item We find that asymmetry in the fractal noise (i.e. asymmetry added on each step of the iterative fractal construction) tends to increase $r$. However, the finite temperature reduces the effects of asymmetry making $r$ more stable against spatial perturbations.
    \item Finally, we consider $T\overline{T}$-deformed 2d CFT --- the integrable non-local deformation, which can be seen as increase of the elementary ``particles'' width (or coupling to dynamical gravity), obtain the dependence for $r$ and comment on it. 
\end{itemize}

The paper is organized as follows. In section 2 we briefly remind how uberholography works for $AdS_3/CFT_2$ and Poincare patch example. In section 3 we discuss higher-dimensional generalization of Cantor-like structures. In section 4 we consider different excited states in uberholography context.

\section{Canonical uberholography}
Let us review the uberholographic proposal discovered in \cite{Pastawski:2016qrs} which has the form of an ultimate recoverability property inherent to holographic duality. The ultimate recoverability means that operators in the bulk can be recovered for zero measure sets on the boundary. In \cite{Pastawski:2016qrs}, this was demonstrated for the Cantor set example and in \cite{Bao:2022vxc} extended on the Sierpinski fractal. As we deal with the holographic correspondence, the cornerstone of all discussion is the Ryu-Takayanagi (RT) formula, relating the minimal surface area spanned on some subregion $A$ on the AdS boundary and hanging into the bulk, to the entanglement entropy of the subregion $A$ in the boundary CFT
\be 
S(A)=\frac{{\it Area}(A)}{4 G}
\ee 
where $G$ is Newton constant. The holographic dual of 2d CFT defined on the line is the metric of the $AdS_3$ Poincare patch, which has the form
\be 
ds^2=\frac{L_{AdS}^2}{z^2}\left(-dt^2+dx^2+dz^2 \right),
\ee 
and the main example of the boundary subregion $A$ to study here is a Cantor set --- a recursively constructed fractal object of zero measure. 

Ryu-Takayanagi surfaces in $AdS_3$ are just geodesics in the Poincare patch, namely, the semicircles spanned on $A$. If $A$ is a single interval of length $\ell$, one can obtain the entanglement entropy of $A$ via the RT formula, namely
\be \label{eq:Sgr}
S(\ell)= \frac{c}{3} \log \frac{\ell}{\varepsilon},
\ee 
where the central charge $c$ is identified with relation $L_{AdS}/G$ as $c=3/2\cdot L_{AdS}/G$ and here $\varepsilon$ is the cut-off. 
The Cantor set is a special set which has a number of properties which is out of the scope of standard intuition, namely it has ``dimensionality'' less than one (in different senses) --- its Hausdorff measure equal to $\log 2/\log 3$, and Lebesgue measure is zero. Cantor set is constructed recursively by poking ``holes'' of certain size in the interval of interest. The main ``building brick'' necessary to understand the entanglement features of the Cantor set is simply two equivalent disjoint intervals, each of length $\ell_1$ and $\ell_2$, and for simplicity we begin with $\ell_1=\ell_2=\frac{r}{2} R$ separated by distance $h=(1-r) R$, such that a total length of the system ``intervals + hole'' is $R$. It is well-known that in a general system of two disjoint intervals of lengths $\ell_{1,2}$ there are two competing configurations of RT surfaces --- disjoint one with the entropy given by
\be 
S_{disj}=\frac{c}{3}\log \frac{\ell_1}{\varepsilon} +\frac{c}{6} \log \frac{\ell_2}{\varepsilon}
\ee 
and the one with ``connected'' topology 
\be 
S_{conn}=\frac{c}{3} \log \left(\ell_1+\ell_2+h\right)+\frac{c}{3}\log h - \frac{2 c}{3} \log \varepsilon.
\ee 
 We are interested in the RT surfaces transition point defined by the minimality condition, i.e. when connected RT topology equals to the disconnected one, $S_{conn}=S_{dis}$. Phase transition of this type  for different models has been considered in \cite{Bernamonti:2012xv,Sircar:2016old}. The solution of this equation for ``building brick'' described above and with $\ell_1=\ell_2$ has the form
\be 
r=2\left(\sqrt{2} - 1 \right ).
\ee

Now let us consider the total system and make symmetric holes in each step of the iteration. The iterations terminate at some length $a$ which can be associated with the natural cut-off or lattice spacing. After $m$ iterations, we have that
$$
a=\left(\frac{r}{2}\right)^{m}|R|,
$$
and the region we are left with (let us call it $R_{\min }$) has $2^{m}$ components of the length $a$ each. Following the proposal of \cite{Pastawski:2016qrs},  the estimate of the distance of the logical operator (sub)algebra $\mathcal{A}_{X}$ associated with the bulk region $X$ is given as follows. The distance $d\left(\mathcal{A}_{X}\right)$ is defined as the minimal size boundary region $R$ which is not correctable with respect to $X$ and the following estimate has been proposed in \cite{Pastawski:2016qrs}
\be 
d\left(\mathcal{A}_{X}\right) \leq \frac{\left|R_{\min }\right|}{a}=2^{m}=\left(\frac{|R|}{a}\right)^{\alpha},
\ee
where 
\be 
\alpha=\frac{\log 2}{\log (2 / r)}=\frac{1}{\log _{2}(\sqrt{2}+1)} \approx 0.786,
\ee
which implies that the distance for any logical subalgebra is bounded by $d\left(\mathcal{A}_{X}\right) \leq n^{\alpha}$, where $n$ could be considered as the number of boundary ``sites'' \cite{Pastawski:2016qrs}.

\section{Higher-dimensional generalization}
\subsection*{Straight slicing}
Now let us consider $AdS_{d+1}/CFT_d$ duality, which relates higher-dimensional gravity and $d-$dimensional boundary field theory. The metric of the $AdS_{d+1}$ Poincare patch has the form
\be 
ds^2=\frac{L_{AdS}}{z^2}(-dt^2+dz^2+d\bar x^2),
\ee 
thus the dual theory is defined on a $d$-dimensional plane. The simplest generalization of the construction described in the previous section to the arbitrary dimension case is to slice $(d-1)$-dimensional constant time sections of the boundary into ``Cantor lasagna'' --- infinitely ``long'' spatial slices with the Cantor set spatial organization in preferred direction. The holographic entanglement entropy of such a single infinitely long slice with an extent of the size $\ell$ in some spatial direction (let us call it $x_1$ for definiteness) is well-known and has the form
\be \label{eq:EE-ddim}
S(\ell)=\frac{1}{4 G^{d+1}}\left[\frac{2 L_{AdS}^{d-1}}{d-2}\left(\frac{\ell_{\perp}}{\varepsilon}\right)^{d-2}-\frac{2^{d-1} \pi^{\frac{d-1}{2}} L_{AdS}^{d-1}}{d-2}\left(\frac{\Gamma\left(\frac{d}{2 d-2}\right)}{\Gamma\left(\frac{1}{2 d-2}\right)}\right)^{d-1}\left(\frac{\ell_{\perp}}{\ell}\right)^{d-2}\right],
\ee 
where again, $\varepsilon$ is the cut-off and $\ell_{\perp}$ is the diverging size of the stripe in the spatial directions complementary to $x_1$. Again, following the logic as in the case of one spatial direction in the boundary system and solving the equation $S_{dis}=S_{conn}$ for the entanglement of the basic element given by \eqref{eq:EE-ddim}, we obtain that now the scrambling length is given by the solution of the equation
\be 
2^d r^2 \left(\frac{L_{\text{AdS}}}{r R}\right){}^d-2 (r-1)^2 \left(\frac{L_{\text{AdS}}}{R-r
   R}\right){}^d-2 \left(\frac{L_{\text{AdS}}}{R}\right){}^d=0.
\ee 
For a particular $d$, this equation has analytic solutions $r$, namely
\bea
&&d=3:\,\, r=3-\sqrt{5}\approx 0.763932,\\
&&d=4:\,\, r=\sqrt{3}-1\approx 0.732051.
\eea 
The dependence of $r$ on $d$ is monotonous, and it is curios to notice, that for large $d \rightarrow \infty$ one can show (at least numerically), that $r$ exactly converges to $r=2/3$. For a generalized Cantor sets, $r=2/3$ delimits the generalized Cantor sets with zero Lebesgue measure and the one with positive measure
\be
d\rightarrow \infty : \,\,\,\, r=2/3.
\ee 
This means that the ``Cantor slicing'' corresponds exactly to infinite number of the dimension $d$ and limits the transition point for the entanglement wedge.

\subsection*{Annular slicing}
Now let us consider another Cantor-like slicing of the plane (to be more precise, the bulk is given by $AdS_4$, with the boundary at constant times being a two-dimensional plane). Instead of straight higher-dimensional belts, now we organize the set under consideration by poking ``holes'' in the form of annular regions. The entanglement entropy of the annular region with two radii $R_1$ and $R_2$ was studied in \cite{Fonda:2014cca} and is given by
$$
S(R_1,R_2)=\frac{c}{6}\left(\frac{2 \pi R_{1}}{\delta}+\frac{2 \pi R_{2}}{\delta}-\frac{4 \pi}{\sqrt{2 \kappa^{2}-1}}\left(\mathbb{E}\left(\kappa^{2}\right)-\left(1-\kappa^{2}\right) \mathbb{K}\left(\kappa^{2}\right)\right)+...\right),
$$
where $\kappa$ is a constant determined by the ratio $R_{2} / R_{1}$
$$
\log \frac{R_{1}}{R_{2}}=2 \kappa \sqrt{\frac{1-2 \kappa^{2}}{\kappa^{2}-1}}\left(\mathbb{K}\left(\kappa^{2}\right)-\Pi\left(1-\kappa^{2} \mid \kappa^{2}\right)\right).
$$
This entanglement is given by the RT surface of the hemi-torus form spanned on the annulus. There is a second phase described by a disconnected solution, which dominates for large $R_2/R_2$. Since we are interested in small size elementary annular regions, we are not interested in this solution. Another expression for annular regions has been obtained in \cite{Han:2019scu} via the entanglement contour proposal
\be 
S(R_1,R_2)=\frac{c}{6}\left(\frac{2 \pi R_{2}}{\epsilon_{2}}+\frac{2 \pi R_{1}}{\epsilon_{1}}-4 \pi \frac{\left(R_{2}^{2}+R_{1}^{2}\right)}{\left(R_{2}^{2}-R_{1}^{2}\right)}+...\right),
\ee
where now $\varepsilon_{1,2}$ are slightly different cut-offs, which leads us to a different formula. Since this expression is much more simple and tractable, let us use it first to estimate $r$. We are interested in thin annular regions, i.e. small $R_2-R_1$. Taking the sizes in terms of $R$ and $r$ as in the previous calculations, it is straightforward to obtain an equation on $r$ for small $R$
\be
\frac{4 \pi  d \left(r^2-6 r+4\right)}{(r-1) r R}+\frac{2 \pi  \left(r^2-6 r+4\right)}{(r-1) r}=0
\ee 
and it has the same solution as for the ``straight'' slicing considered above in the text
\be 
r=3-\sqrt{5},
\ee 
which confirms the universality of this quantity.

\section{Excited states: finite temperature, $T \overline{T}$}
As we saw in the previous examples, in general, the distance $d\left(\mathcal{A}_{X}\right)$ is defined by $r$, and since we are interested in the reconstruction of regions with the fractal dimension, only the UV behaviour corresponding to small basic fractal building blocks matters. For the higher-dimensional ``straight-belt'' slicing, the result for $r$ is fixed solely by the dimension, leading to the universal estimate for $d\left(\mathcal{A}_{X}\right)$ on all scales, while the radial slicing in $d=3$ gives us the same $r$ only in the UV limit.

\subsection*{Finite temperature}
\begin{figure}
    \centering
	\includegraphics[width=0.6\textwidth]{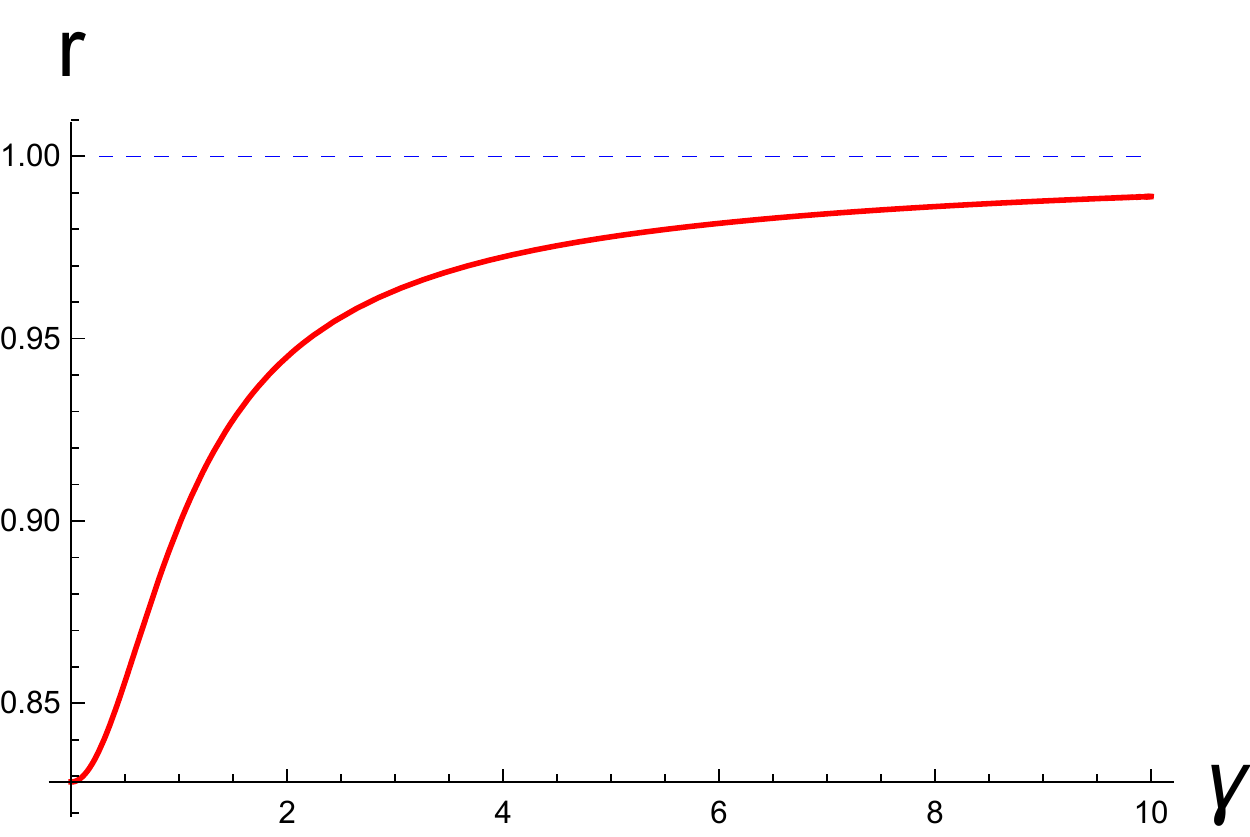}
	\caption{The dependence of $r$ on $\gamma$ given by \eqref{eq:rdepT}.}
	\label{fig:rbtz}
\end{figure}

It is interesting to consider some tractable examples of excited states in the similar context. One of the simplest generalization of the Poincare metric is the BTZ black hole
\be 
d s^{2}=\frac{L^{2}}{z^{2}}\left(-f(z) d t^{2}+\frac{d z^{2}}{f(z)}+d x^{2}\right), \quad f(z)=1-z^{2} / z_{h}^{2},\,\,\,\,T=\frac{1}{2\pi z_h},
\ee 
dual to 2d CFT at finite temperature $T$ fixed by horizon location. The entanglement entropy for a single interval is given by the corresponding RT surface, leading to the entanglement entropy
\be 
S(\ell)=\frac{c}{3} \log \left(\frac{\sinh \left(\pi T \ell \right)}{\pi T \varepsilon}\right).
\ee 
For the considered previously spatial organization of the boundary region, the parameter $r$ now has the form explicitly depending on scales through $R$ and $T$, namely
\be \label{eq:rdepT}
r=\frac{2 \coth ^{-1}\left(\coth (\pi  \gamma )+\sqrt{2} \text{csch}(\pi  \gamma )\right)}{\pi  \gamma },
\ee 
where $\gamma=R\cdot T$. Now the distance $d\left(\mathcal{A}_{X}\right)$ is controlled not only by the size $R$ but also by the temperature. In principle, according to this estimate taking large enough temperature, one can obtain the corresponding corrections to the distance $d\left(\mathcal{A}_{X}\right)$. The dependence of $r$ on $\gamma$ is presented in Fig.\ref{fig:rbtz}, indicating that for large enough temperature (and small enough $R$) the distance converges to the estimate $d\left(\mathcal{A}_{X}\right)<n$.

\begin{figure}
    \centering
	\includegraphics[width=0.55\textwidth]{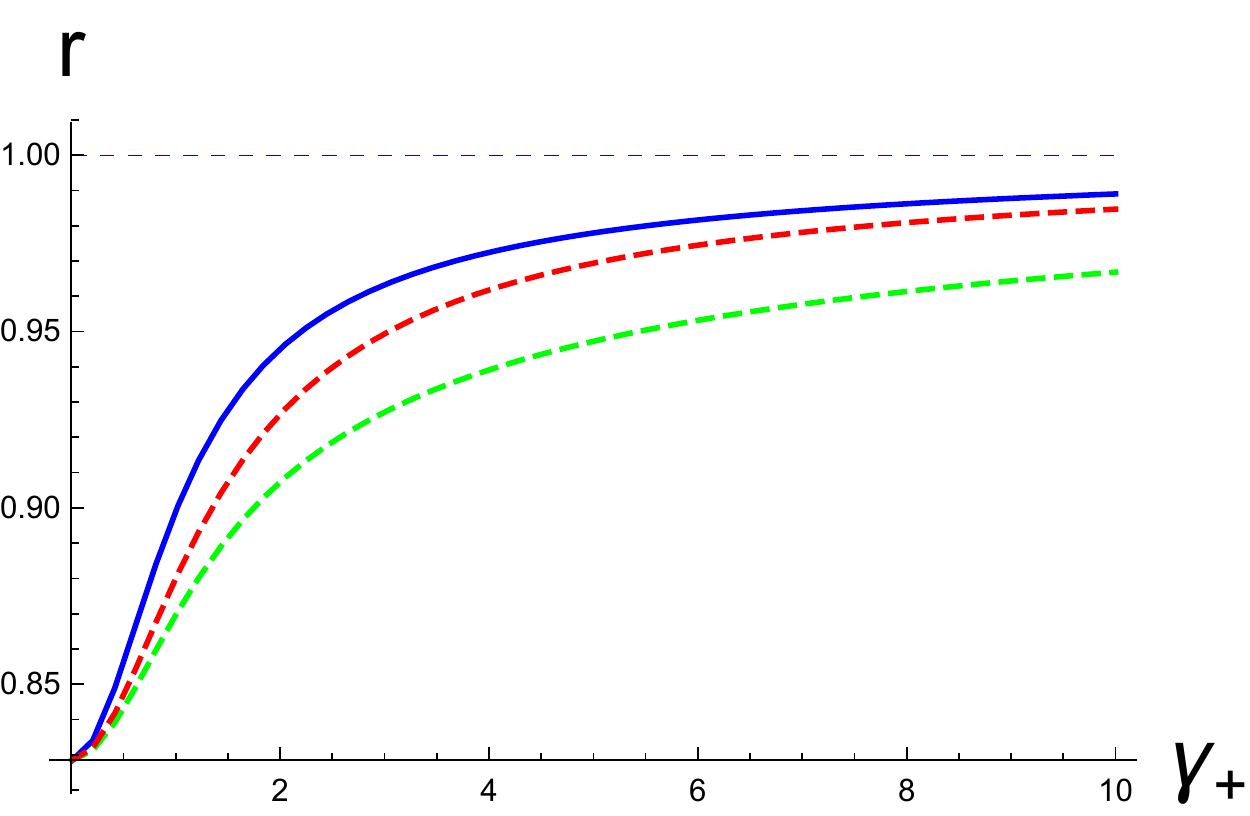}
	\caption{The dependence of $r$ on $\gamma_+$ for different $\gamma_-$ (green dashed curve corresponds to $\gamma_- \rightarrow 0$, red one to $\gamma_-=\gamma_+/2$. Blue solid curve corresponds to the non-rotating black hole.}
	\label{fig:rrot}
\end{figure}

Now consider how the rotating and extremal black holes fit into this construction. Assuming that $z=1/r$, the canonical form of the rotating BTZ black hole metric has the form
\be 
d s^{2}=-\frac{\left(r^{2}-r_{+}^{2}\right)\left(r^{2}-r_{-}^{2}\right)}{r^{2}} d t^{2}+\frac{r^{2}}{\left(r^{2}-r_{+}^{2}\right)\left(r^{2}-r_{-}^{2}\right)} d r^{2}+r^{2}\left(d \phi-\frac{r_{+} r_{-}}{2 r^{2}} d t\right)^{2},
\ee 
and the effect of rotation introduces two different temperatures $T_{\pm}$ for the left and right moving modes of dual CFT (apart the ordinary black hole temperature)
\be 
T_{+}=\frac{r_{+}+r_{-}}{2 \pi}, \quad T_{-}=\frac{r_{+}-r_{-}}{2 \pi},\,\,\,\,\,\,\,\,T=\frac{r_{+}^{2}-r_{-}^{2}}{2 \pi r_{+}}.
\ee 
The single interval entanglement entropy of the dual theory now depends on the temperatures $T_\pm$ and explicitly, has the form
\be 
S(\ell)=\frac{c}{6} \log \left(\frac{\sinh \left(T_+\pi \ell\right)}{T_+\pi \epsilon}\right) +\frac{c}{6} \log \left(\frac{\sinh \left(T_-\pi \ell\right)}{T_-\pi \epsilon}\right),\,\,\,\,
\ee 
which lead us to the equation determining $r$
\be
\frac{\sinh \left(\pi  \gamma_-\right) \sinh \left(\pi  \gamma_+\right) \sinh \left(\pi  (1-r )
   \gamma_-\right)  \sinh \left(\pi    (1-r ) \gamma_+\right) }{\text{sinh}^2\left(\frac{1}{2} \pi  r  \gamma_-\right) \text{csch}^2\left(\frac{1}{2} \pi  r  \gamma_+\right)}=1.
\ee 
In Fig.\ref{fig:rrot}, we plot the dependence of $r$ on $\gamma_+$ for different $\gamma_-$ and  lower $\gamma_-$ (i.e. closer to extremality) decrease the growth of $r$.

\subsection*{Asymmetric regions: thermal stabilization and smoothness property?}
\begin{figure}
    \centering
	\includegraphics[width=0.45\textwidth]{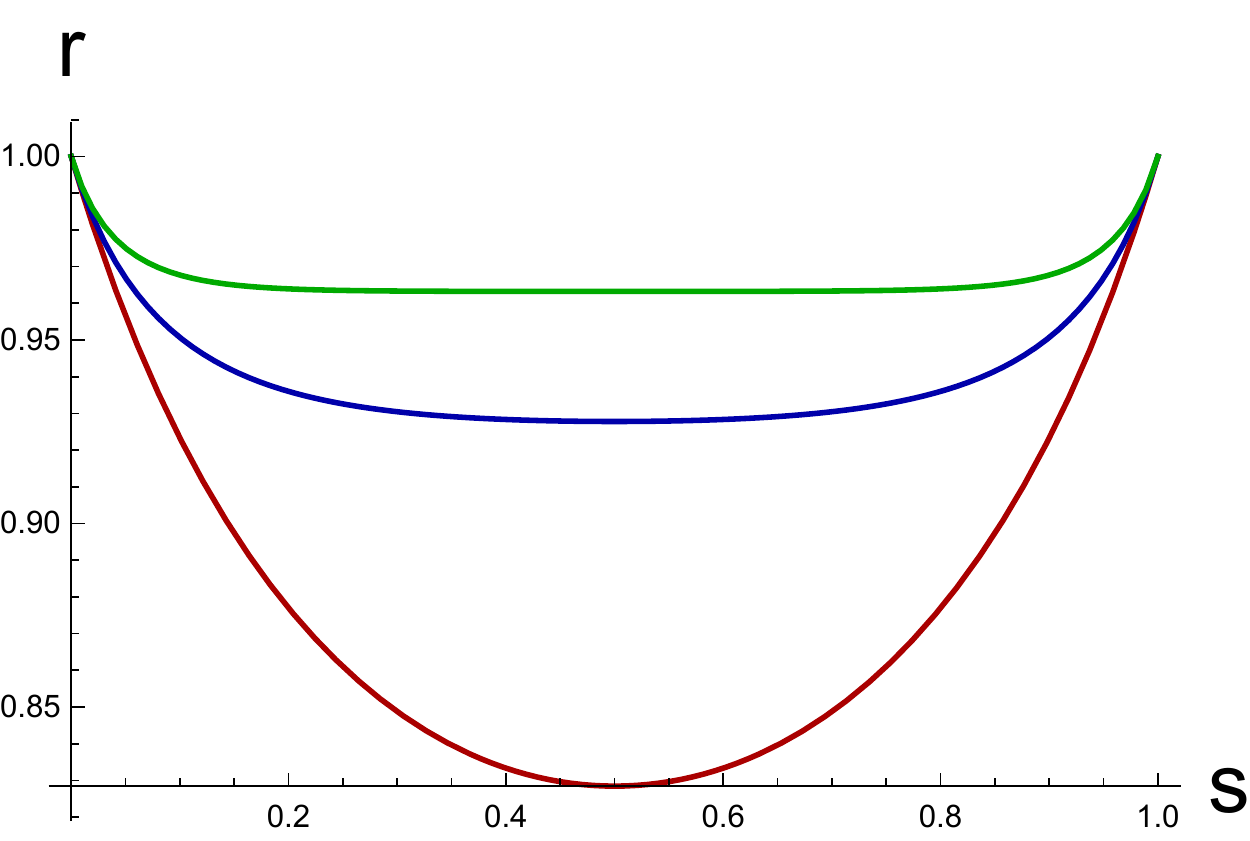}
	\caption{The dependence of $r$ on the asymmetry parameter $s$ for different values of $\gamma$ (red curve corresponds to zero temperature, blue one to $\gamma=1.5$ and green one to $\gamma=3$).}
	\label{fig:rst}
\end{figure}

Now let us turn to the question, what happens if we slightly deviate from the Cantor set and how the temperature affects such construction in the context of uberholography. The Cantor set is constructed by iterative withdraw exactly of the middle third of each subinterval. In fact, even the small variations of the withdraw procedure parameters can lead us to different topological characteristics of the final set. So it is interesting to consider what happens after some changes of the procedure  in the uberholographic context. First, let us consider an asymmetric partition of the interval with the length $R$ obtained by choice 
\be 
\ell_1=(1-s) R r,\,\,\,\,\ell_2=s R r,\,\,\,\, h=(1-r)R,
\ee 
i.e on the each step of iteration, we obtain the partition of the system into inequal parts separated by the distance $(1-r) R$. For the ground state, i.e. when the entropy is given by \eqref{eq:Sgr}, we obtain the solution for $r$ corresponding to the entanglement wedge phase transition explicitly in the form
\be \label{eq:asymm}
r=\frac{2}{\sqrt{1-4 (s-1) s}+1}.
\ee 
It is straightforward to see that $r$ has the minimum at $s=1/2$, i.e. precisely for the Cantor set. The generalization of equation \eqref{eq:asymm} to the finite temperature case is straightforward, although not leading to an analytic formula which may be written in some readable way. In Fig.\ref{fig:rst}, we plot the dependence of $r$ on $s$ for different temperatures. Although this does not lead to the generalization of $d\left(\mathcal{A}_{X}\right)$ and $\alpha$ (for arbitrary $s$) in a straightforward manner, one can try to estimate the qualitative features solely by the behaviour of $r(s)$.
From this plot, one can see that the dependence of $r$ on $s$ becomes more stable, i.e. a large enough temperature deviation from $s=1/2$ does not lead to a significant change in $r$.

\subsection*{$T\overline{T}$ - deformation}
\begin{figure}
    \centering
	\includegraphics[width=0.45\textwidth]{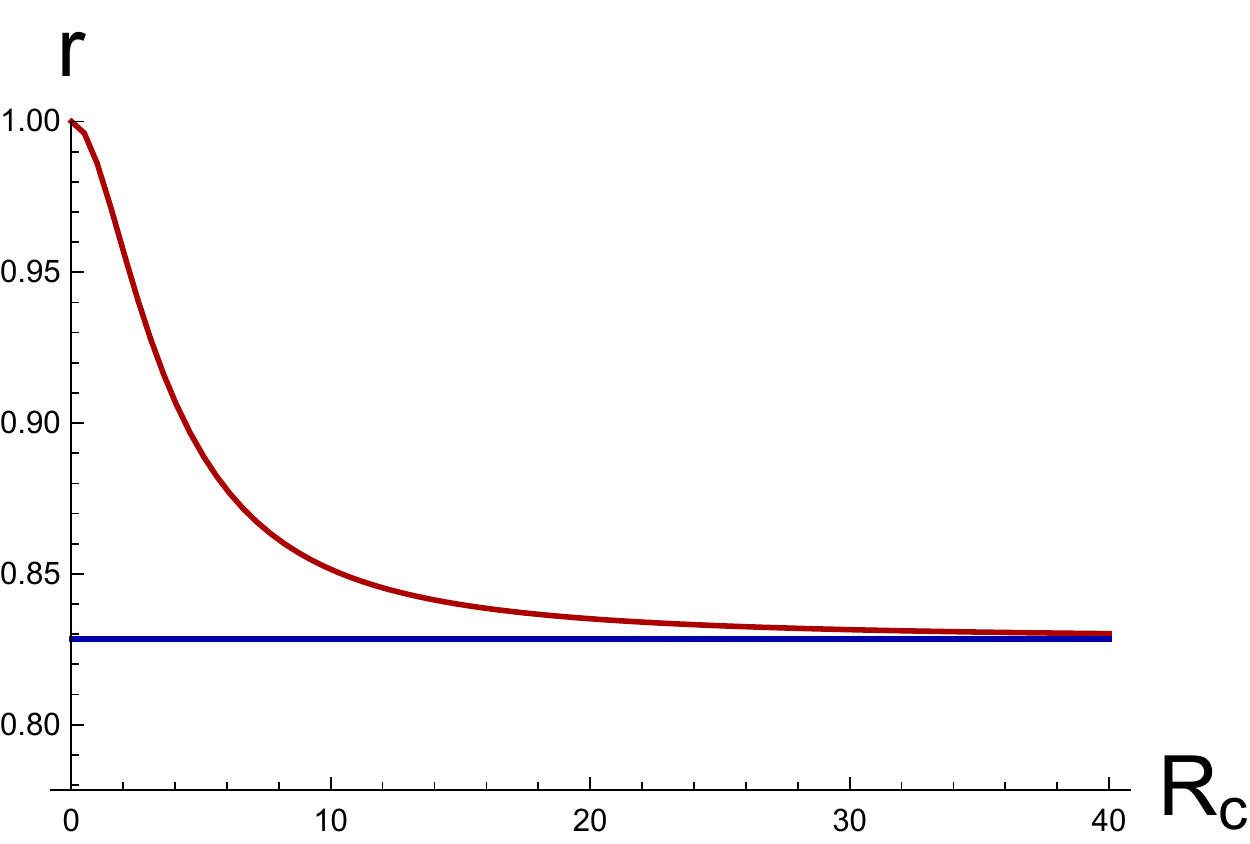}
	\caption{The dependence of $r$ on the $R_c=R/z_c$ for $T\overline{T}$-deformed theory}.
	\label{fig:TTbar}
\end{figure}
Finally, let us consider the special kind of non-local exactly solvable quantum field theories --- $T \overline{T}$ deformed 2d CFT \cite{Zamolodchikov:2004ce,Cavaglia:2016oda,Smirnov:2016lqw,Conti:2018tca}. The bulk dual of $T \overline{T}$ deformation of 2d CFT is given by the situating the boundary theory at a radial cut-off of the bulk at some finite $z_c$ \cite{McGough:2016lol}. This deformation admits different interpretations, for example it could be considered as the coupling of field theory to JT gravity \cite{Dubovsky:2017cnj,Dubovsky:2012wk} or some kind of dynamical coordinate transformation applied to the field. Also, one can consider the deformation as the broadening of the fundamental particles ``width'' (extending/reduction of the phase space or wavefunction support in the non-relativistic case)\cite{Cardy:2020olv,Medenjak:2020ppv}.
The computation of the entanglement entropy\footnote{The holographic entanglement entropy in $T \overline{T}$ deformed theories has been considered in \cite{Donnelly:2018bef,Chen:2018eqk,Park:2018snf}. } in different ways has been shown to be consistent with the prescription given by the cut-off proposal.

The entanglement entropy of the interval with the length $\ell$ is now given by
\be 
S(\ell)= \log \left(\sqrt{\frac{\ell_{c}^2}{4 }+1}+\frac{\ell_c}{2} \right),\,\,\,\,\ell_c=\frac{\ell}{z_c}
\ee 
which vanishes when $\ell \rightarrow 0$ in contrast to the logarithmic divergence in the undeformed theory. At leading nonvanishing order, the solution of the equation defining $r$ for $R_c=R/z_c \rightarrow 0$ is given by
\be 
T\overline{T}\text{-deformed 2d CFT:}\,\,\,\,r=1,\,\,\,\, R_c\rightarrow 0
\ee 
due to linearly decreasing entanglement entropy $S(\ell)\approx \ell/(2 z_c)$. From Fig.\ref{fig:TTbar}, one can observe that the coefficient $r$ being equal $r=1$ for $R\rightarrow 0$ is size dependent and for large $\ell_c$, it decreases to the ordinary $r=2(\sqrt{2}-1)$ corresponding to 2d CFT vacuum. The interpretation in terms of the phase space deformation (or broadening the width of fundamental particles) seems to be consistent with this change of $r$. For small $\ell_c$, the fine structure of the fractal cannot be taken into account due to the finite size of particles. It would be interesting to clarify the effect of $T \overline{T}$ deformations in explicit quantum error correcting codes and systems using objects of ``zero size'' (like in the Gotesman-Preskill-Kitaev protocol which uses infinite sequence of Dirac delta functions to encode a qubit).

\section*{Acknowledgements}
This work was funded by Russian Federation represented by the Ministry of Science and Higher Education
(grant number 075-15-2020-788).

\end{document}